\documentclass[
  pra,
  reprint,
  superscriptaddress,
  nofootinbib,
  amsmath,
  amssymb,
  floatfix
]{revtex4-2}

\usepackage{braket}
\usepackage{bm}

\usepackage{graphicx}
\usepackage{booktabs}
\usepackage{array}
\usepackage{multirow}
\usepackage{dcolumn}

\usepackage{xcolor}

\usepackage{xurl}

\newcommand{\Diss}{\mathcal{D}}
\newcommand{\Tr}{\operatorname{Tr}}
\newcommand{\Wg}{\mathcal{W}_{\mathrm g}}
\newcommand{\Wl}{\mathcal{W}_{\mathrm l}}

\usepackage[
  colorlinks=true,
  linkcolor=blue,
  citecolor=blue,
  urlcolor=blue
]{hyperref}
\begin{document}

\title{Engineering Coherence and Ergotropy through Spatial Arrangement  of Environmental Channel in a Two-Qubit Quantum Battery}

\author{Gayatree Swain}
\email{g.swain@iitg.ac.in}
\affiliation{Department of Physics, Indian Institute of Technology Guwahati, Guwahati, India}

\date{\today}

\begin{abstract}
We investigate how the spatial arrangement of environment-channel influences the dynamics of coherence and work extraction in a two-qubit quantum battery with resonant qubits coupled through an XY-exchange interaction. Qubit B is attached to a finite-temperature thermal bath, while RTN noise is applied either to qubit A or to qubit B, depending on whether it is in a separated or co-located configuration, respectively. Two complementary product states are considered to initialise the coherence on different qubits. The dynamics are evaluated using local $l_{1}$-norm coherence, global ergotropy and the global-local ergotropy gap. The results suggest that a stronger exchange interaction enhances the redistribution of coherence without necessarily improving retention of extractable work. For the parameter investigated, the greater global ergotropy is retained in the case of the co-located configuration, especially when the initially coherent qubit is not directly affected by the RTN. However, the configuration provides the largest global-local ergotropy gap, which varies with the exchange interaction, initial state and thermal coupling. Suppression of the gap at large $J$ is typically observed in the case of stronger thermalisation, although enhancement of intermediate-coupling may occur. These results establish that total extractable work, local coherence and collective work advantage quantify the distinct aspects of quantum battery performances. The initial distribution of coherence and the spatial arrangement thus provide complementary mechanisms for regulating locally and collectively extractable work in interacting open quantum batteries.
\end{abstract}
\maketitle
\section{Introduction}

Quantum thermodynamics examines how heat, energy and work are processed in microscopic systems, where correlations, coherence and quantum measurement can affect thermodynamic behavior\cite{Goold2016,Campaioli2024}. In this context, there is development of a finite-dimensional quantum system in the form of quantum batteries, which can store and extract energy through control operation \cite{Campaioli2019,Campaioli2024}. Previous studies demonstrated that, compared to independent control of individual battery cells, collective quantum operation can improve charging power and extractable work \cite{Campaioli2017,Alicki2013}. These outcomes motivate to investigate more on quantum resources and control strategies, which can improve energy storage at microscopic scales.\\
Ergotropy provides an operational measure of quantum battery performance and displays the maximum extractable work from a state through cyclic unitary transformations \cite{Allahverdyan2004}. In contrast to average stored energy, ergotropy excludes the passive part of energy that cannot be converted into work by unitary control\cite{Pusz1978,Allahverdyan2004}. Extractable work is influenced by correlation and coherence, even without changing total energy\cite{Alicki2013,Cakmak2020}. In particular, global operations may access work that remains inaccessible to independent local transformation \cite{Alicki2013}. This distinction naturally motivates one to compare locally accessible ergotropy with global ergotropy, whose differences characterise the work advantage offered by collective discharge operations. A quantum battery is, in reality, unavoidably affected by its environment. Stored energy and coherence are generally degraded by dissipation and dephasing, but their effects is not universally detrimental. Under suitable Markovian and non-Markovian conditions, noise has been shown to enhance transient ergortropy and accelerate charging \cite{Ghosh2021,Kamin2020,Liu2024}. These results show that the sources of loss should not only be regarded as environmental effects, as their structure, placement and strength can influence how quantum resources and energy are distributed within a battery.\\
Random telegraph noise RTN is a widely used model, especially for low-frequency classical fluctuations in solid-state qubits. It is demonstrated by a stochastic field switching between values, $+\mu$ and $-\mu$, at a rate $\nu$. RTN can produce either rapidly damped dynamics or slowly modulated oscillations. With partial coherence revivals, which depend on the ratio between the noise amplitude and the switching rate of RTN \cite{Paladino2002,LoFranco2012}. Such outcomes make RTN  a useful model for investigating the redistribution of coherence and memory effects in coupled quantum systems. Nevertheless, comparatively less explored of how the location of RTN relative to thermal energy-exchange channels influences coherence and work extraction in an interacting quantum battery\cite{Ghosh2021,Liu2024}. In the presence of appreciable RTN, the system spectrum changes with the instantaneous noise value; this motivates applying a branch-dependent global dissipator, which is separate from two conditioned Hamiltonians\cite{Davies1974,Breuer2002}.\\
We investigate two resonant qubits coupled by an XY- exchange interaction, which is also coupled to RTN and a finite-temperature thermal bath. The qubit B is always connected to the reservoir, whereas RTN acts either on qubit B in the co-located configuration or on qubit A for separated configuration. The initial states $\ket{0}_{A}\ket{+}_{B}$ and  $\ket{+}_{A}\ket{0}_{B}$ are chosen to examine the effects of placing the initial coherence on different subsystems. The dynamics are characterised using local coherence, global ergotropy and the global-local ergotropy gap as a function of $J$, $\kappa$ and environment placement. Our results demonstrate that stringer exchange strength promotes redistribution of coherence without necessary improvement of work retention. Moreover, different channel configurations may be required to maximise global ergotropy; therefore, initial coherence placement and environment geometry may provide distinct means of controlling work accessibility in coupled quantum batteries. 

\section{Model}
\subsection{Hamiltonian for the coupled-qubit system}
 consider two resonant coupled qubits, denoted by $A$ and $B$, through an interaction involving XY exchange. Setting $\hbar=1$, the total Hamiltonian is given by 
\begin{equation}
 H_{total}=H_{i}+H_{\mathrm{int}},
 \label{eq:Htotal}
\end{equation}
where,
\begin{equation}
H_{i} = \frac{\omega}{2}( \sigma_{ZA} + \sigma_{ZB})
\end{equation}

Initially, the two qubits were uncoupled and evolved independently under the free Hamiltonian $H_{i}$. After we introduce the interaction Hamiltonian XY exchange, it allows quantum information and energy to be exchanged between qubits A and B.\\
Under this interaction, the total excitation number is preserved while transferring an excitation between qubits \cite{Bose2003,Du2025}.\\
\begin{equation}
H_{\mathrm{int}}
=
J(\sigma_{A X}\sigma_{BX}+ \sigma_{A Y}\sigma_{BY})
\end{equation}
Here, $\omega$ is the energy splitting of each qubit and  $J$ is the coupling strength. 
The interaction terms couple the single-excitation basis states$\ket{01}$ and $\ket{10}$. \\
When $J\neq 0$, these states are not the eigenstates of an interacting Hamiltonian. Diagonalisation of the interacting Hamiltonian yields the coupled qubit energy eigenstates.

\begin{equation}
 \ket{\psi_\pm}=\frac{\ket{01}\pm\ket{10}}{\sqrt{2}},
\end{equation}
with corresponding  energies $E_{\psi_\pm}=\pm2J$. The remaining eigenstates $\ket{00}$ and $\ket{11}$ of the full Hamiltonian, with energies, are $E_{00}=+\omega$ and $E_{11}=-\omega$.\\
For the regime of energy parameters $0<J<\omega/2$, the eigenstates are ordered from highest to lowest energy as $\ket{00}$, $\ket{\psi_+}$, $\ket{\psi_-}$, and $\ket{11}$.

\begin{figure}[tbp]
    \centering

    % First row
    \includegraphics[width=\columnwidth]
    {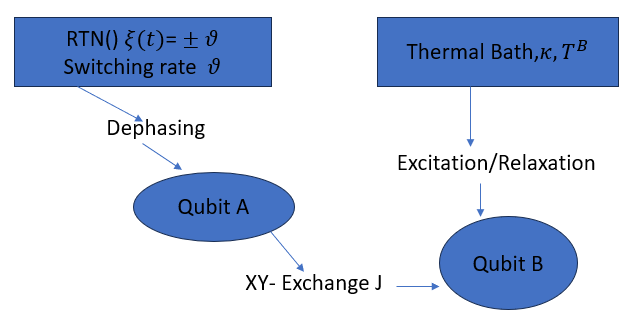}

    \par\smallskip
    \textbf{(a)}
    \par\medskip

    % Second row
    \includegraphics[width=\columnwidth]
    {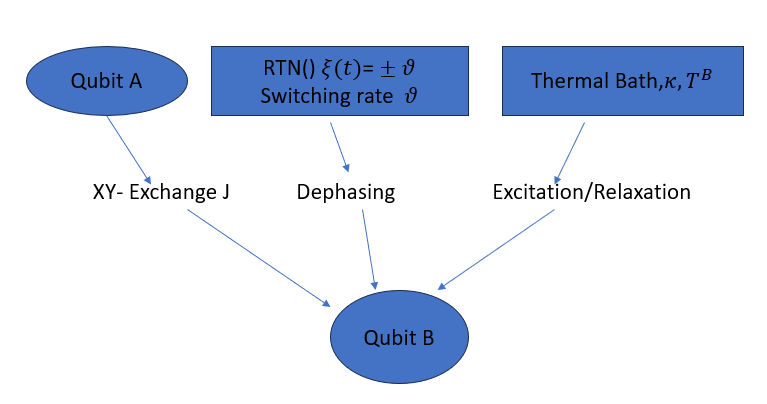}

    \par\smallskip
    \textbf{(b)}

    \caption{Schematic representation of the two environmental-channel
    configurations: (a) separated configuration, in which RTN acts on
    qubit \(A\) while the thermal reservoir is coupled to qubit \(B\);
    and (b) co-located configuration, in which both RTN and the thermal
    reservoir act on qubit \(B\). In both configurations, qubits \(A\)
    and \(B\) are coupled through the XY exchange interaction.}
    \label{fig:channel-configurations}
\end{figure}
\subsection{ channel-placement configurations}
To study how the location of initial coherence influences the open-system dynamics and extractable work, we consider two complementary initial product states. In each state, one qubit is in an energy eigenstate, while the other is in a coherence superposition.  
\begin{equation}
 \ket{\Psi_1(0)}=\ket{0}_A\ket{+}_B,
 \qquad
 \ket{\Psi_2(0)}=\ket{+}_A\ket{0}_B,
 \label{eq:initialstates}
\end{equation}
where $\ket{+}=(\ket{0}+\ket{1})/\sqrt{2}$.This choice allows us to place the initial local coherence either on qubit A  or on qubit B, while keeping the total initial-state structure otherwise comparable. In the first state, local coherence is located on qubit $B$, while the second contains local coherence initially on qubit $A$.
The two open-system configurations are shown schematically in Figs.~\ref{fig:channel-configurations}(a) and
\ref{fig:channel-configurations}(b). In both configurations, The thermal bath is always connected to qubit $B$. The thermal bath can exchange energy with $B$, causing thermal excitation and relaxation.  In the separated configuration, the thermal bath acts on $B$, while the RTN dephasing acts on $A$, separating dephasing from the thermal energy exchange. In the co-located configuration, both RTN dephasing and the thermal bath act on  $B$, co-locating the two channels.

\subsection{ Thermal-Bath and Random-Telegraph Dephasing}
Here, we frame the model with a non-Markovian dephasing channel using (RTN) Random-telegraph noise. The RTN field function is given by  $\xi(t)$, which randomly takes either $+\mu$ or $-\mu$ and switches between these two values at a rate $\nu$. $\mu$ denotes the strength of the random phase fluctuation, while $\nu$ is the switching rate, indicating how rapidly the environment changes between its two configurations \cite{Daffer2004,Bergli2009,Paladino2014}. To describe the dynamics of RTN, we take two auxiliary conditional density matrices, $\rho_{+}$ and $\rho_{-}$. These conditioned matrices are generally individually unnormalized. $\rho_{+}$ and $\rho_{-}$ conditioned on the noise begins $+\mu$ and $-\mu$, respectively. 
By summing the physical density matrices of the two-qubit system.

\begin{equation}
 \rho(t)=\rho_+(t)+\rho_-(t),
 \qquad 
 \rho_\pm(0)=\frac{\rho(0)}{2}.
 \label{eq:physicalrho}
\end{equation}
In both configurations, the thermal bath is coupled to qubit B. Qubit B can relax and transfer energy to the bath through the lowering operator $\sigma^{B}_{-}$, and it can absorb energy from a finite-temperature bath through the raising operator $\sigma^{B}_{+}$ where $\kappa$ determine the strength of the system-bath coupling and for a thermal bosonic bath at temperature T, the mean thermal occupation number, at transition frequency $\omega$ is, $n_{th} = [\exp{\frac{\omega}{k_{B}T}}-1]^{-1}$
at zero temperature, $ n _ {th} = 0$; the bath induces only the spontaneous downward transition, thereby removing excitation from the two-qubit system\cite{Breuer2002}. At finite temperature, the bath couples to qubit B; the internal interaction of XY exchange between the two qubits leads to redistribution of energy and coherence.

The thermal bath on $B$ is modeled by
\begin{equation}
 \mathcal{L}_{\mathrm{th}}^B[\rho]=
 \kappa(n_{\mathrm{th}}+1)\Diss[\sigma_-^B]\rho+
 \kappa n_{\mathrm{th}}\Diss[\sigma_+^B]\rho,
 \label{eq:thermal}
\end{equation}
This is the standard weak-coupling thermal Lindblad master equation for a two-level system interacting with the bosonic reservoir. \cite{Breuer2002}.
where 
$\Diss[L]$ denotes the Lindblad dissipator. The term proportional to $\Diss[\sigma^{B}_{-}]$ describes relaxation of qubit B, whereas the term proportional to $\Diss[\sigma^{B}_{+}]$indicates the thermal excitation. $\sigma^{B}_{-}$ = $\ket{1}_{B}\bra{0}$ and $\sigma^{B}_{+}$ = $\ket{0}_{B}\bra{1}$ where, the qubit B takes from the higher energy state $\ket{0}_{B}$ to the lower energy state $\ket{1}_{B}$. It represents one relaxation event, corresponding to
\begin{equation}
 \Diss[L]\rho=L\rho L^\dagger-\frac{1}{2}\{L^\dagger L,\rho\}.
\end{equation}
Equation~\eqref{eq:thermal} is the local weak-coupling thermal dissipator. RTN provides phase noise, whereas the thermal bath exchanges energy with the system.

\subsection{Model 1: Separated configuration }
The stochastic field can take the value $\xi(t) = \pm\mu$, the system evolve via two conditional density operators, $\rho^{1}_{+}$ and $\rho^{1}_{-}(t)$. The dynamic for the weaker-coupling case are governed by,

\begin{align}
 \dot{\rho}^{(1)}_+={}&-i[H+vZ_A,\rho^{(1)}_+]
 +\mathcal{L}_{\mathrm{th}}^B[\rho^{(1)}_+]
 +\nu(\rho^{(1)}_- -\rho^{(1)}_+),\\
 \dot{\rho}^{(1)}_-={}&-i[H-vZ_A,\rho^{(1)}_-]
 +\mathcal{L}_{\mathrm{th}}^B[\rho^{(1)}_-]
 +\nu(\rho^{(1)}_+ -\rho^{(1)}_-).
 \label{eq:model1}
\end{align}
The Hamiltonian, describe the evolution associated with the two possible value of the RTN field. The terms proportional to $\nu$ accounts for switching between these noise branches. 

\subsection{Model 2: Co-located Configuration}

For Model~2,  the corresponding dynamics are governed by, 
\begin{align}
 \dot{\rho}^{(2)}_+={}&-i[H+vZ_B,\rho^{(2)}_+]
 +\mathcal{L}_{\mathrm{th}}^B[\rho^{(2)}_+]
 +\nu(\rho^{(2)}_- -\rho^{(2)}_+)\label{eq:model11},\\
 \dot{\rho}^{(2)}_-={}&-i[H-vZ_B,\rho^{(2)}_-]
 +\mathcal{L}_{\mathrm{th}}^B[\rho^{(2)}_-]
 +\nu(\rho^{(2)}_+ -\rho^{(2)}_-).
 \label{eq:model12}
\end{align}
The preceding equations ~\eqref{eq:model11} and ~\eqref{eq:model12} constitute RTN conditioned local master equation because the thermal jump operator $\sigma_{\pm}^{B}$ acts on the bare energy levels of qubit B \cite{Breuer2002,Levy2014,Scali2021}. The XY-coupling is fully incorporated into the unitary part of the system dynamics, but it is not included when constructing the dissipative jump operator.

\subsection{Global-master equation}
In the strong-coupling regime, the bath can resolve the energy splitting generated by an XY interaction, and the thermal bath induces transitions between the dressed eigenstates of the complete Hamiltonian of the system\cite{Davies1974,Breuer2002,Levy2014,Scali2021}. Furthermore, the RTN amplitude need not be perturbatively small; each RTN-conditioned Hamiltonian should be treated separately.
for model $p = 1,2$ and RTN branches  $q = \pm1$,, the hamiltonian is defined as,
\begin{equation}
    H_{q}^{(p)} = H \pm q\mu\sigma_{Xp}
\end{equation}
where $X_{1}= A$  in the model-1
where $X_{2}= B$  in the model-2
Its spectral decomposition is;
\begin{equation}
    H_{q}^{(p)} = \sum_{E} E_{q}^{p}\chi_{E,q}^{p}
    \end{equation}
where $E_{p}^{q}$ and $\chi_{E,q}^{p}$ represents, eigen-energies and eigenvectors respectively,
In both cases, qubit-B is directly coupled to the thermal bath through the system operator $S_{B}$, which is taken to be $S_{B} = \sigma_{-}^{B} + \sigma_{+}^{B}$. This operator decomposes into components associated with the positive Bohr frequencies $f$ of $H _{q}^{(p)}$ gives,
\begin{equation}
A_{q}^{(p)}(f) = \sum_{E{'}-E =f } \pi_{E,q}^{(p)} S_{B} \pi_{E{'},q}^{(p)}
\label{eq:global thermal}
\end{equation}, 
Here, the $A_{q}^{(p)}(f)$ operator induces a downward transition releasing energy $f = E_{'} -E > 0 $ into the reservoir and $A^{ (p)\dagger }_{q}(f)$corresponding represents, thermally induced upward transition\cite{Davies1974,Breuer2002}.  \\
Using equation~\eqref{eq:global thermal},the global thermal dissipator is therefore,

\begin{equation}
\begin{aligned}
\mathcal{L}_{G,q}^{(p)}
&= \sum_{f>0}\gamma(f)\left[n_{B}(f)+1\right]
\Diss[A_{q}^{(p)}(f)]\rho \\
&\quad + \sum_{f>0}\gamma(f)n_{B}(f)
\Diss[A_{q}^{(p)\dagger}(f)]\rho .
\end{aligned}
\end{equation}
 $\gamma(f)$ denotes bath-induced transition rate at frequency $f$ and 
 
 \begin{equation}
   n_{B}(f) = \frac{1}{\exp(\beta f)-1}
 \end{equation}
 
This is mean thermal occupation of the reservoir at that frequency  \cite{Davies1974,Breuer2002}. Combining conditioned unitary evolution, global thermal dissipation, and symmetric RTN switching gives the global master equation \cite{Bergli2009,Paladino2014} as,
 \begin{align}
  \dot{\rho}^{(p)}_+={}&-i[H_{q}^{(p)},\rho^{(p)}_q]
  +\mathcal{L}_{\mathrm{G,q}}^{(p)}[\rho^{(p)}_q]
  +\nu(\rho^{(p)}_{-q} -\rho^{(p)}_{+q}) 
 \end{align}
 For each channel configuration, the physical density operator is obtained by summing the two RTN branches.
 \begin{equation}
 \rho^{p}(t) = \rho^{(p)}_{+}(t) + \rho^{(p)}_{-}(t)
 \end{equation}
 with the initial conditions,
 \begin{equation}
     \rho_{\pm}^{(p)}(0)  = \frac{1 }{2}\rho(0)
 \end{equation}

\section{Coherence and work measure}

The reduced density matrices of qubits A and B are obtained by taking the partial trace of the complete two-qubit density operator over the other qubit,\\
 $\rho_A=\Tr_B[\rho(t)]$ and $\rho_B=\Tr_A[\rho(t)]$. 
 The coherence stored locally in each subsystem is quantified in the bare energy basis using the $l_{1}$-norm \cite{Baumgratz2014}. For a single qubit, this measure is given,  

 \begin{figure}[tbp]
    \centering

    \includegraphics[width=7cm,height=5cm]
    {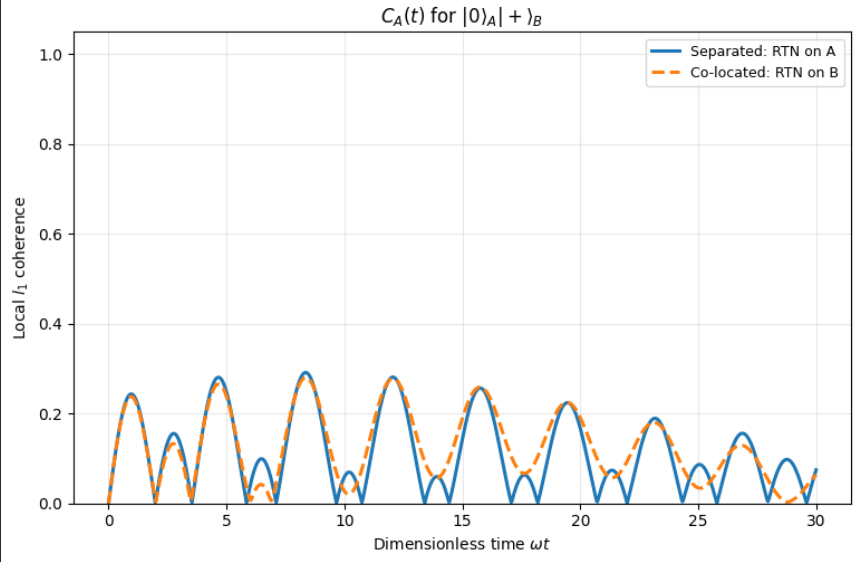}

    \par\smallskip
    \textbf{(a)}
    \par\medskip

    \includegraphics[width=7cm,height=5cm]
    {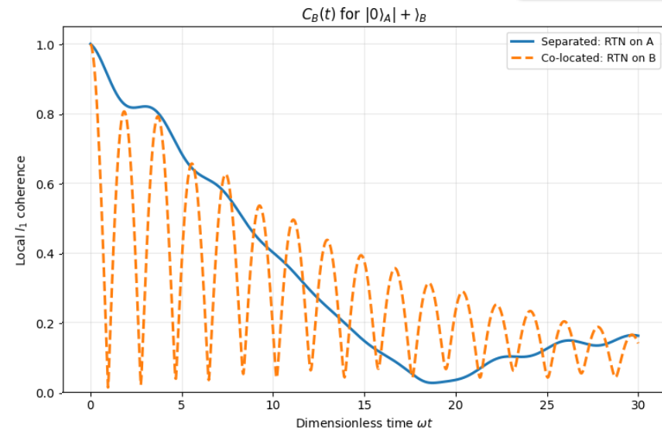}

    \par\smallskip
    \textbf{(b)}

    \caption{Local-coherence dynamics for \(J=0.2\omega\):
    (a) coherence \(C_A(t)\) of qubit \(A\), and
    (b) coherence \(C_B(t)\) of qubit \(B\).
    The remaining parameters are
    \(\kappa=0.05\omega\), \(\mu=0.8\omega\),
    \(\nu=0.02\omega\), and \(\beta=\ln(6)/\omega\).}
    \label{fig:local-coherence-J02}
\end{figure}

\begin{figure}[tbp]
    \centering

    \includegraphics[width=7cm,height=5cm]
    {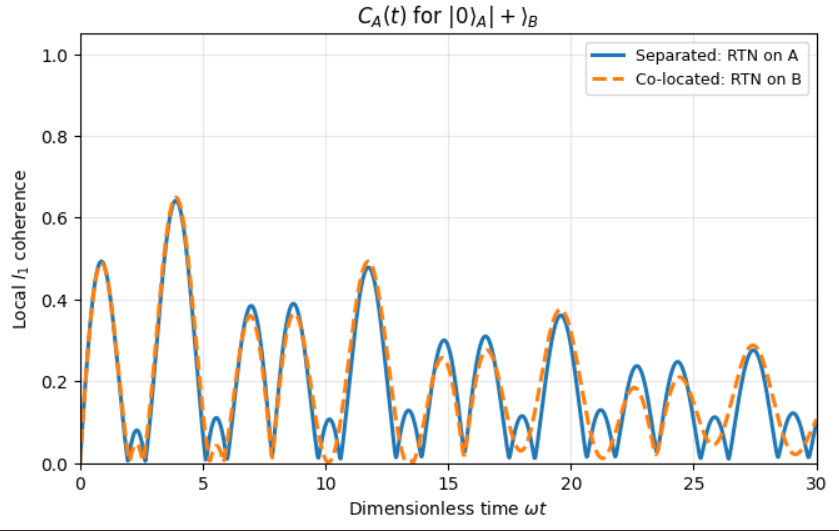}

    \par\smallskip
    \textbf{(a)}
    \par\medskip

    \includegraphics[width=7cm,height=5cm]
    {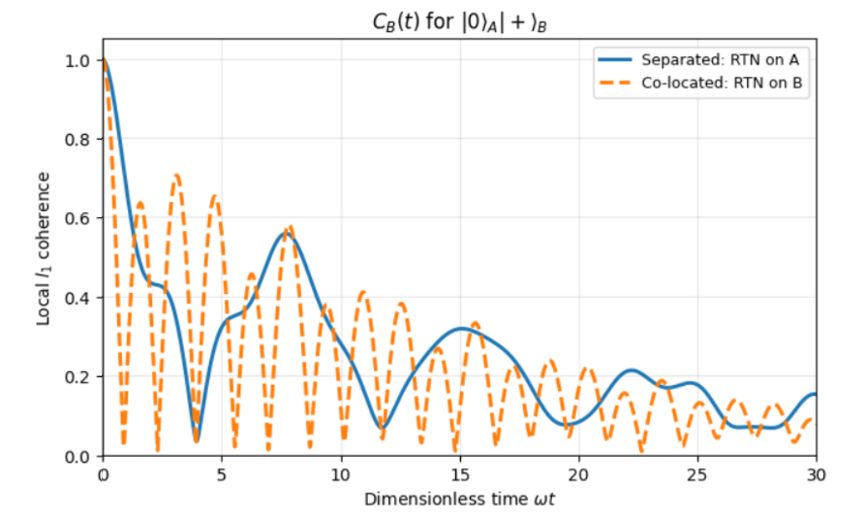}

    \par\smallskip
    \textbf{(b)}

    \caption{Local-coherence dynamics for \(J=0.45\omega\):
    (a) coherence \(C_A(t)\) of qubit \(A\), and
    (b) coherence \(C_B(t)\) of qubit \(B\).
    The remaining parameters are
    \(\kappa=0.05\omega\), \(\mu=0.8\omega\),
    \(\nu=0.02\omega\), and \(\beta=\ln(6)/\omega\).}
    \label{fig:local-coherence-J045}
\end{figure}

Here,$C_{A}(t)$ and $C_{B}(t)$ represents the coherence between the local energy eigenstates of qubits A and B. For the initial state $\ket{0}_{A}\ket{+}_{B}$, qubit A is initially incoherent while the coherence of qubit B is maximal; hence $C_{B}(0) = 1$ and $C_{A}(0) = 0$. Figure~\ref{fig:local-coherence-J02}  for $J=0.2\omega$, and Figure~\ref{fig:local-coherence-J045} for $J= 0.45\omega$ presents $C_{A}(t)$ and $C_{B}(t)$ as a function of the dimensionless time $\omega t$ for the separated and co-located configuration. These results illustrate how this process is influenced by RTN and the thermal environment and demonstrate how the XY interaction redistributes the initial coherence between qubits.
We employ the concept of ergotropy to evaluate the work-storage capability of the two-qubit system. Global ergotropy represents the maximum amount of energy that can be extracted from the state $\rho(t)$ through an arbitrary unitary transformation of the two-qubit system~\cite{Allahverdyan2004}. It is defined as:
\begin{equation}
 \mathcal{W}=\Tr[\rho(t)H]-\Tr[\rho_{\mathrm{p}}(t)H].
\end{equation}
Here, $\rho_{p}(t)$ represents the passive states corresponding to $\rho(t)$, while the eigenvalues are arranged in decreasing order and assigned to the energy eigenstates of $H$, ordered from the lowest to the highest energy. Through unitary operations, no further energy can be extracted from a passive state\cite{Pusz1978,Allahverdyan2004}, so the difference between $\rho(t)$ and $\rho_{p}(t)$ gives the maximum work available under unrestricted joint control\cite{Allahverdyan2004}.
Let the eigenvalues of the density operator be ordered as $S_{1}> S_{2}>.. $ and let the energy eigenvalues of $H $ satisfy $E_{1}< E_{2}<....$ and the passive state energy is,
$E_{p}(t) = \sum_{i}S_{i}E_{i}$. To compare global and local control while retaining the interaction during the discharge stroke, we define the locally accessible work as,

\begin{equation}
\begin{aligned}
\mathcal{W}_{\mathrm l}(t)
={}&
\operatorname{Tr}\!\left[\rho(t)H\right]
\\
&-
\min_{U_{\mathrm l}=U_A\otimes U_B}
\operatorname{Tr}\!\left[
U_{\mathrm l}\rho(t)U_{\mathrm l}^{\dagger}H
\right].
\end{aligned}
\label{eq:local-ergotropy}
\end{equation}

This restricted work-extraction functional corresponds to an optimization over product unitary operations while interacting; the Hamiltonian is retained \cite{Castellano2025}.
The global-local work difference is
\begin{equation}
 \Delta\mathcal{W}_{\mathrm{GL}}(t)=\Wg(t)-\Wl(t).
 \label{eq:gap}
\end{equation}
We also track the interaction energy as,
\begin{equation}
 E_{\mathrm{int}}(t)=\Tr[\rho(t)H_{\mathrm{int}}].
 \end{equation}
 
\begin{figure}[tbp]
    \centering

    \includegraphics[width=7cm,height=5cm]
    {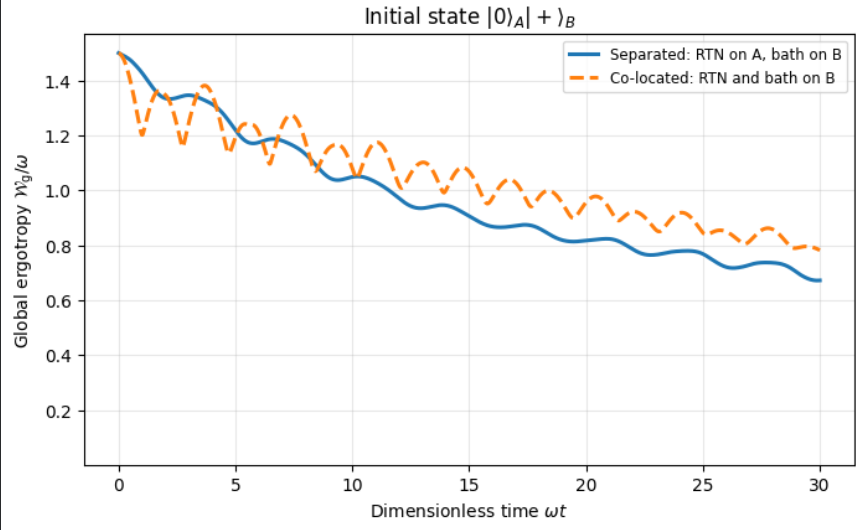}

    \par\smallskip
    \textbf{(a)}
    \par\medskip

    \includegraphics[width=7cm,height=5cm]
    {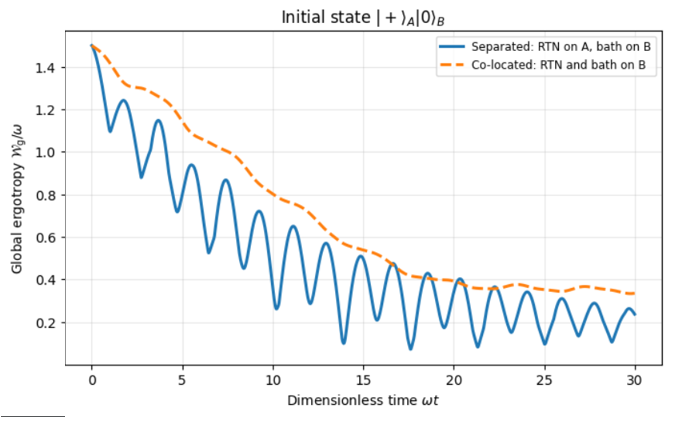}

    \par\smallskip
    \textbf{(b)}

    \caption{Global-ergotropy dynamics for \(J=0.2\omega\):
    (a) initial state \(\ket{0}_A\ket{+}_B\), and
    (b) initial state \(\ket{+}_A\ket{0}_B\).
    The remaining parameters are
    \(\kappa=0.05\omega\), \(\mu=0.8\omega\),
    \(\nu=0.02\omega\), and \(\beta=\ln(6)/\omega\).}
    \label{fig:global-ergotropy-J02}
\end{figure}

\begin{figure}[tbp]
    \centering

    \includegraphics[width=7cm,height=5cm]
    {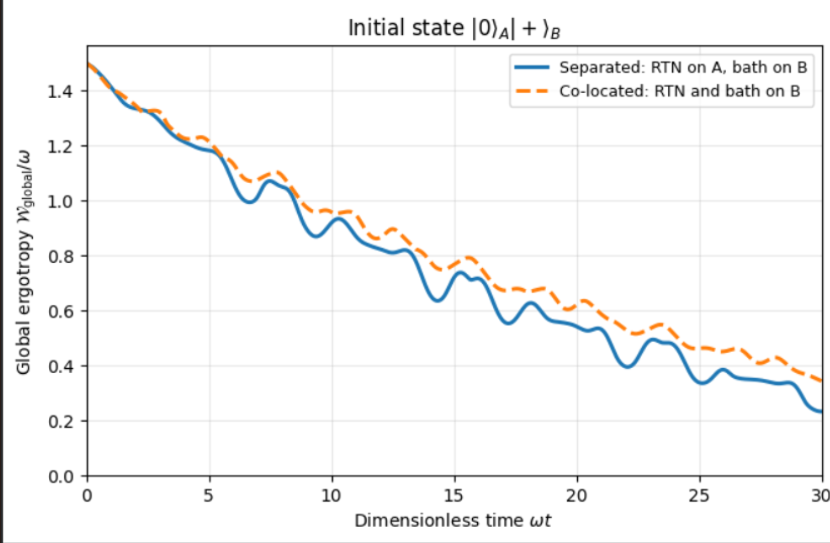}

    \par\smallskip
    \textbf{(a)}
    \par\medskip

    \includegraphics[width=7cm,height=5cm]
    {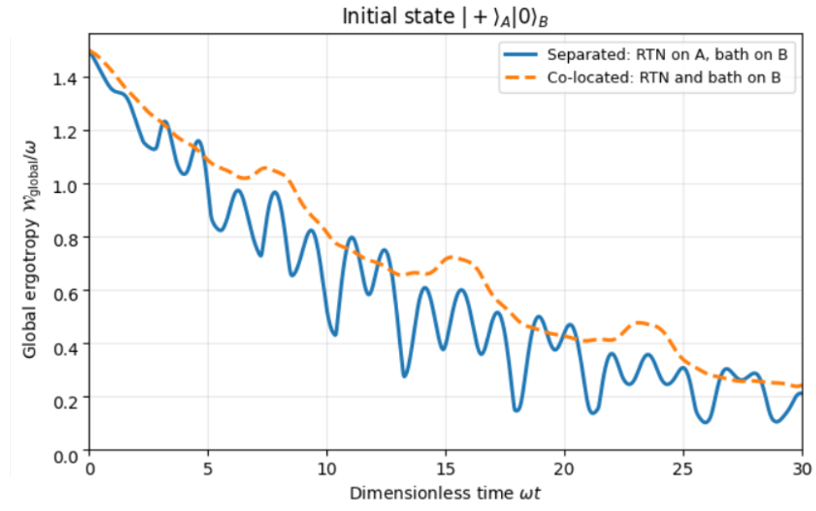}

    \par\smallskip
    \textbf{(b)}

    \caption{Global-ergotropy dynamics for \(J=0.45\omega\):
    (a) initial state \(\ket{0}_A\ket{+}_B\), and
    (b) initial state \(\ket{+}_A\ket{0}_B\).
    The remaining parameters are
    \(\kappa=0.05\omega\), \(\mu=0.8\omega\),
    \(\nu=0.02\omega\), and \(\beta=\ln(6)/\omega\).}
    \label{fig:global-ergotropy-J045}
\end{figure}

Figure~\ref{fig:global-ergotropy-J02} for $ J=0.2\omega$ and Figure~\ref{fig:global-ergotropy-J045} for $ J=0.45\omega$ show the time evolution of the global ergotropy, plotted against the dimensionless time $\omega t$ for model-1 and model-2. We select $ J=0.45 \omega $and $ J=0.2\omega $ as representative lower- and higher-exchange cases within $0<J< \omega/2 $. These values generate well-separated exchange timescales and substantially different dressed transition spectra while preserving the same eigenstate ordering. Both cases are treated using the RTN-branch resolved global dissipator so that the observed differences can be attributed to the variation of $J$. 
\section{Results}
\subsection{Local coherence dynamics}

Figure~\ref{fig:local-coherence-J02}  for $J=0.2\omega$, and Figure~\ref{fig:local-coherence-J045} for $J= 0.45\omega$ displays the local $l_{1}$-norm coherence $C_{A}(t)$ and $C_{B}(t)$ for the initial state $\ket{0}_{A}\ket{+}_{B}$ at $\omega t =0$, all coherence resides in qubit B, giving $C_{A}(0) = 0$ and $C_{B}(0)= 1$. the XY interaction redistribute this coherence between the qubits. Consequently, oscillatory coherence develops in qubit A, whereas qubit B undergoes simultaneous reduction and revival of coherence. In the lower-exchange case, $ J=0.2\omega$ is shown in Figure~\ref{fig:local-coherence-J02} for $J=0.2\omega$, the coherence generated in qubit A does not exceed the value of approximately 0.29. The separated and co-located curves for$C_{A}(t)$ almost overlap at early times and differ only moderately at later times. This means that coherence transfer to qubit A is governed mainly by the XY-exchange interaction and the dependence on the RTN source is weaker for this selected parameter. In contrast, the coherence of qubit B is strongly affected by channel placement. In the separated configuration, where RTN acts on qubit A, $C_{B}(t)$ decreases relatively smoothly, with broad oscillatory variation produced by XY-exchange coupling. Whereas in the co-located case, RTN and the thermal reservoir are co-located on B, coherence $C_{B}(t)$ undergoes frequent collapses followed by partial recoveries. The slow-noise character from $\frac{\nu}{\mu}= 0.025$ places the RTN in the slow-switching regime, for which the fields remain near either $+\mu$ or $-\mu$ for relatively long periods. For the higher-exchange case, $ J = 0.45\omega$, the display in Figure~\ref{fig:local-coherence-J045} shows the significant enhancement of the coherence that transfers to qubit A, reaching a peak value of approximately 0.65. This increased peak amplitude indicates that increasing the XY-coupling promotes a more substantial redistribution of the initial local coherence. In addition, oscillations are more rapid with a more complex modulation structure, reflecting the increased exchange rate and dressed-state transition frequencies of the interacting system.\\
For the higher-exchange case, the separated and co-located curves for $C_{A}(t)$ remain close over much of the evolution and the influence of the principal placement again appears in $C_{B}(t)$. In the co-located configuration, $C_{B}(t)$ develops frequent collapses and revivals. Whereas the separated configuration produces broad damped oscillations. At long times, both arrangements preserve a small amount of coherence, with weak residual oscillations arising from the persistent interplay among the XY exchange, RTN and thermal reservoir. As J increases from 0.2$\omega$ to 0.45$\omega$, there is a rise in the maximum generated coherence in the qubit, from roughly 0.29 to 0.65, accompanied by a faster redistribution of coherence between the qubits. The results therefore suggest that the XY-interaction mainly sets the strength and timescale of coherence transfer. Whereas the position of RTN mainly controls the collapse and revival behaviour of $C_{B}(t)$, since qubit B initially contains the coherence. The results for both \(J=0.2\omega\) and \(J=0.45\omega\) were evaluated using the RTN-branch resolved global thermal dissipator. The comparison between the two cases describes the effect of varying the exchange strength within a consistent open-system framework.

\subsection{Global ergotropy and channel placement}
Figure~\ref{fig:global-ergotropy-J02} for $J = 0.2\omega$ and Figure~\ref{fig:global-ergotropy-J045} for $J= 0.45\omega$ shows the evolution of the global ergotropy for the two initial states,$\ket{0}_{A}\ket{+}_{B}$ and $\ket{+}_{A}\ket{0}_{B}$  under both the separated and co-loacted channel arrangements. the parameters are fixed at $\kappa = 0.05\omega $, $\mu = 0.8\omega $ and $\nu = 0.02 \omega $. Both configuration curves begin at approximately the same global ergotropy, $W_{g}(0) \approx 1.5\omega$, because the environment configurations have identical initial conditions at t = 0 and systems Hamiltonian. The subsequent separation and oscillation indicate that extractable work is associated with both the position of the initial coherence and the relative placement of the thermal bath to RTN. For $J = 0.2\omega $, the initial state $\ket{0}_{A}\ket{+}_{B}$, the evolution of global ergotropy decreases gradually with time. Although separated and co-located arrangements represent similar behaviour during the initial evolution, the later part of the evolution, the co-located configuration generally retains a slightly larger ergotropy at $\omega t = 30$ the co-located curve approaches a value of $0.79\omega$. Whereas, the separated curve retains approximately 0.67$\omega $. The oscillation observed in the co-located curve indicated the evolving interplay among thermal relaxation, coherence excitation exchange and RTN.
For the swapped initial state $\ket{+}_{A}\ket{0}_{B}$ a substantially  different behavior is observed. In a separated configuration, qubit A is applied directly by the RTN channel, which initially carries the coherence, resulting in a comparatively fast reduction of extractable work and strong oscillations. In the co-located configuration, follows smooth envelopes and remain higher through most of the evolution. However, the two curves approach one another at later times. This again shows that separating the initial coherence from the directly noisy qubit can enhance finite-time work retention. \\
At $J= 0.45\omega $ for both initial states, the global ergotropy decreases more strongly. For the initial states  $\ket{0}_{A}\ket{+}_{B}$, the separated and co-located curves follow comparable decay trends, but the co-located configuration maintains a consistently higher ergotropy over most of the evolution. At $\omega t = 30$, the global ergotropy is roughly $0.35\omega$ and $0.23\omega $for the co-located and separated configurations, respectively. The higher-exchange interaction case introduces more visible short-time modulation, reflecting bath-induced transitions among the dressed energy levels of the coupled system.
For a swapped state $\ket{+}_{A}\ket{0}_{B}$, the difference between the two curves remains clearly visible. In the separated configuration, the RTN is directly acting on qubit A having initial coherence, leading to large oscillations with a rapidly decreasing amplitude. In comparison, the co-located configuration generally preserves higher ergotropy and develops a smoother decay envelope, although the two curves approach one another at later times. These observations indicate that ergotropy retention depends on the placement of the RTN channel relative to initial coherence. The Comparison of two interaction strengths shows that $ J = 0.45\omega$ generally reaches a lower value at long times than at $J = 0.2\omega$. stronger exchange interaction does not necessarily improve the preservation of globally extractable work. Studies above state that increasing J promotes faster transfer of energy and coherence between the qubits. It also facilitates transfer to the bath-coupled qubit and modifies the available dressed-state relaxation pathways. The faster ergotropy reduction observed at $J=0.45\omega$ is consistent with this competition. Since both interaction strengths are evaluated using the same RTN-branch resolved global master equation, their comparison describes the influence of J within a consistent dissipative framework.  
For $\ket{0}_{A}\ket{+}_{B}$, the initial coherence resides on qubit B, which always coupled to the thermal bath as a result, the two ergotropy curves remain comparatively close. For the initial state $\ket{+}_{A}\ket{0}_{B}$, the coherence at an early stage is located on qubit A, and RTN is directly on the coherent qubit in a separated configuration. This leads to a much stronger suppression of extractable work and more pronounced oscillations. This shows that the relative advantage of a particular channel placement cannot be explained solely by whether RTN and the bath are separated or co-located; it also depends on the location of initial prepared coherence. The local coherence dynamics qualitatively reflect the behaviour of global ergotropy. For the initial state $\ket{0}_{A}\ket{+}_{B}$, on increasing J develops higher coherence peaks on A, demonstrating stronger coherence redistribution between qubit B and A. Nevertheless, at $J = 0.45\omega$ there is a more pronounced decay of the global ergotropy. Thus, stronger transfer of local coherence does not necessarily lead to improved retention of extractable work. Global ergotropy depends jointly on the complete two-qubit state, including coherence correlation, its energy content and population distribution. In the co-located configuration, the rapid collapses and partial revivals of $C_{B}(t)$ are reflected weakly in the global-ergotropy plot. This shows that RTN driven local-coherence revivals can temporarily restore part of the states energetic activities. However, an increase in local coherence need not generate a comparable increase in global ergotropy, because thermal dissipation can simultaneously decrease the energy of the system and organise its population toward a more passive configuration.

\subsection{Parameter robustness}
\begin{figure*}[t]
    \centering

    % First row
    \includegraphics[width=0.95\textwidth]
    {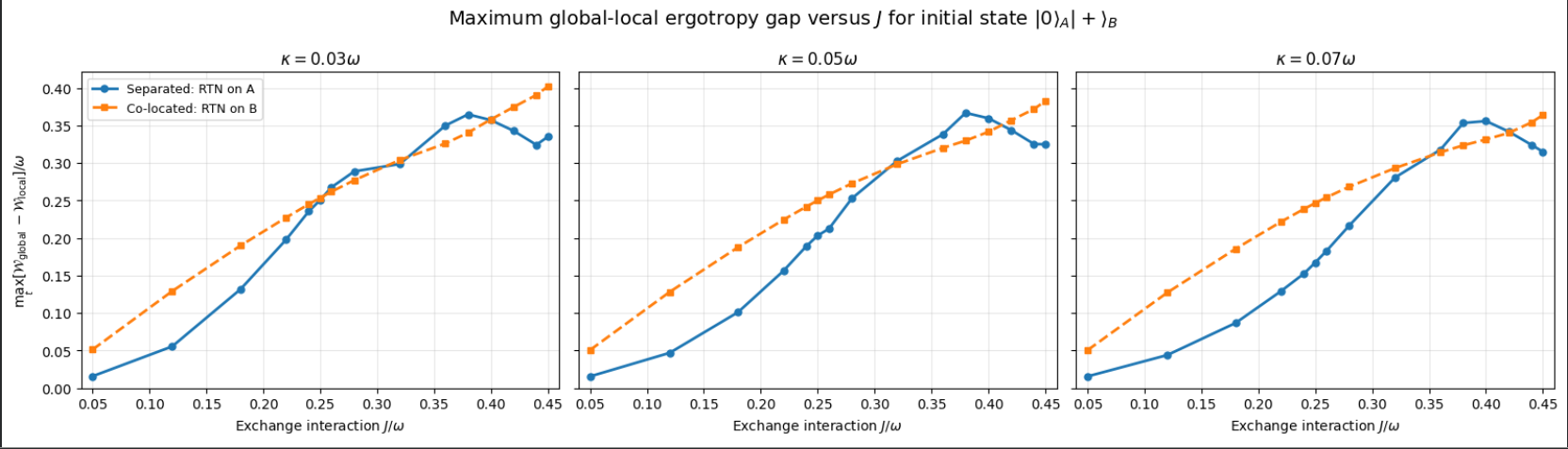}

    \par\smallskip
    \textbf{(a)}
    \par\medskip

    % Second row
    \includegraphics[width=0.95\textwidth]
    {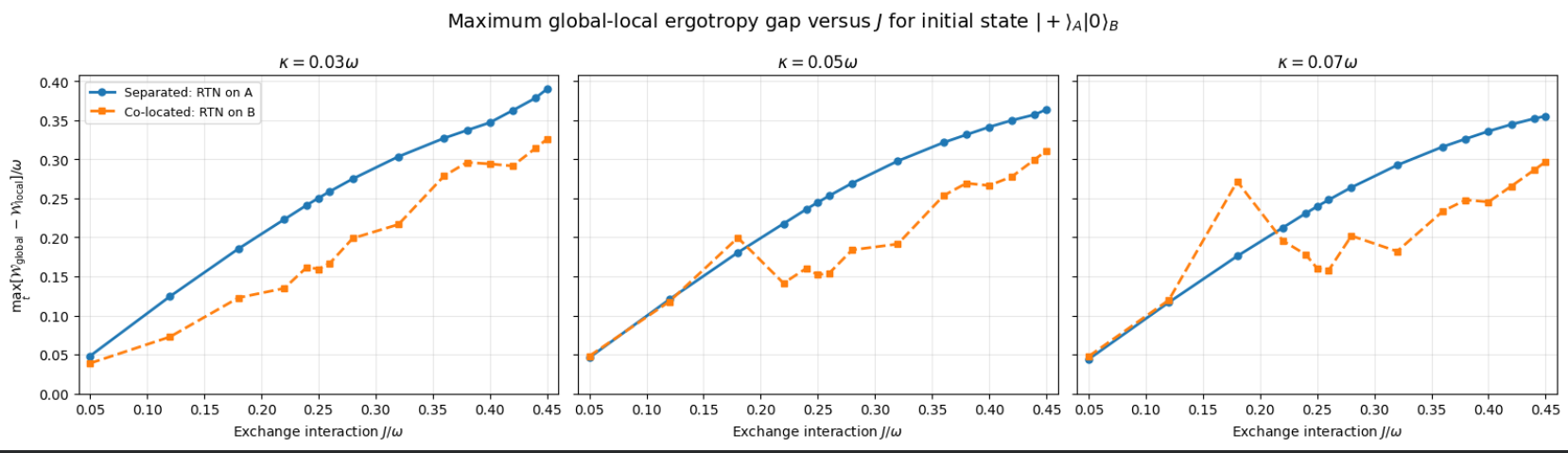}

    \par\smallskip
    \textbf{(b)}

    \caption{Maximum global--local ergotropy gap as a function of
    \(J/\omega\) for (a) the initial state
    \(\ket{0}_A\ket{+}_B\) and (b) the swapped initial state
    \(\ket{+}_A\ket{0}_B\). The three thermal-coupling strengths are
    \(\kappa=0.03\omega\), \(0.05\omega\), and \(0.07\omega\).
    The remaining parameters are \(\mu=0.8\omega\),
    \(\nu=0.02\omega\), and \(\beta=\ln(6)/\omega\).}
    \label{fig:maximum-global-local-gap}
\end{figure*}

Figure~\ref{fig:maximum-global-local-gap} compares the maximum
global--local ergotropy gap for the two initial states and three
thermal-coupling strengths. Demonstrates the maximum difference between global and locally accessible ergotropy. This curve is evaluated as a function of the normalized exchange interaction $\frac{J}{\omega}$ for three values of thermal interaction strength. This quantity measures the maximum work advantages offered by unrestricted collective operations over product-unitary control. A non-zero value denotes the usefulness of joint control, but need not be interpreted directly as a measure of entanglement.For the initial state $\ket{0}_{A}\ket{+}_{B}$.\\
Both the curve shows an increase in the maximum ergortropy gap with J. For the co-located configuration, a larger gap is obtained in the weaker exchange region at intermediate J  and then the two curves approach one another. However, the separated configuration produces a maximum near $J \approx 0.35\omega$, followed by a moderate decline, and the co-located gap continues to rise. Increasing $\kappa$ from $0.03\omega$ to $0.07\omega$ slightly reduces the largest gap and changes the crossover between the two configurations, showing that stronger thermalisation competes with collective work advantage generated by exchange coupling. \\
The swapped initial state $\ket{+}_{A}\ket{0}_{B}$ shows a noticeably different placement observed. The separated configuration shows an almost monotonic increase of $\Delta\mathcal{W}_{max}$ with J and exceeds the co-located values across most of the investigated range. The co-located curves are more non-monotonic, particularly at $\kappa = 0.07\omega$, where intermediate J enhancement is followed by subsequent reduction and recoveries. When initial coherence is stored in qubit A, its direct coupling to RTN in the separated arrangement therefore develops a greater difference between the global and locally accessible work. In comparison, the two initial states demonstrate that the joint-control advantage is determined by the combined influence of $ J$ and $ \kappa$ and the initial position of coherence. The maximum-ergotropy gap provides information complementary to that contained in the global-ergotropy dynamics.\\ 
Although for the initial state  $\ket{+}_{A}\ket{0}_{B}$, the co-located arrangements generally preserve more global-ergotropy, it frequently demonstrates a small difference between global and locally accessible work. Showing that a larger fraction of the retained work is accessible through local control. Conversely, a larger gap in the separated configuration need not indicate larger total work content; instead, it demonstrates that joint two-qubit operations are required to access a greater fraction of remaining work. \\
 Noticeably difference is observed when these results are compared with local-coherence dynamics. Although an increase of J generally strengthens the transfer of coherence between the qubits, its effect on $\Delta\mathcal{W}_{max}$ is influenced by both the channel-placement and the initial state. For characterising different aspects of battery operation, we therefore need three quantities: Local coherence, which describes the redistribution of coherence at the subsystem level. 
Global-ergotropy measures the total amount of extractable work, and  $\Delta\mathcal{W}_{max}$  quantifies the additional work advantage enabled by collective discharge protocols.
\section{Conclusion}
We examined how the spatial arrangement of RTN and a thermal bath affects work extraction and coherence in an XY-coupled two-qubit quantum battery. Two complementary models were considered: a co-located arrangement in which the RTN and thermal bath are located on qubit B, and a separated configuration arranges RTN which acts on qubit A and the thermal bath on qubit B. \\
Comparing the initial states $\ket{0}_{A}\ket{+}_{B}$ and $\ket{+}_{A}\ket{0}_{B}$, we identified the combined role of the location of initial coherence, channel placement, exchange coupling J and thermal coupling $\kappa $. The Local-coherence evolution with time shows that the XY-exchange interaction transfers coherence between the qubits.\\
By increasing J, the magnitude and rate of this transfer enhances, which is demonstrated by the large coherence developed by the initially incoherent qubit. The qubit that initially carries the coherence is primarily affected by the placement of the channel. When that qubit is directly acted upon by RTN, its coherence exhibits stronger collapses and partial revivals, especially in the slow-switching region considered here. The separated and co-located cases can therefore develops noticeable different local dynamics, even when their Hamiltonian and environment strengths are otherwise the same.\\
The Global-ergotropy outcome displays stronger coherence transfer but need not necessarily improve work retention. Although increasing J enhances faster distribution of coherence and energy, this state that its show direct useful resources toward the qubit coupled to the bath and activates additional dressed-state relaxation pathways.
Consequently, the region of strong coupling generally exhibits a faster decreasing of global ergotropy. The co-located configuration preserves greater extractable work over much of the evolution for the studies of parameters, with the advantage being especially pronounced for $\ket{+}_{A}\ket{0}_{B}$, where the RTN does not directly act on the initial coherence. This shows that the performance of the battery depends not only on whether environmental channels are co-located or separated, but also on their relative placement to the initially coherent subsystem.\\
The maximum global-local ergotropy gap reveals further differences between the total amount of stored work and its accessibility under restricted control. For the initial state $\ket{0}_{A}\ket{+}_{B}$, the preferred arrangement changes with J: separated is advantageous in the intermediate coupling region, whereas co-located becomes competitive at weak and strong coupling regions. \\
For the swapped initial state, the separated configuration develops the greater collective-control advantage. Over most of the investigated range, there is a suppression of the strong-coupling gap with increasing $ \kappa$, although non-monotonic curves are shown in the intermediate region of J, when the arrangements are located. These trends show that thermalisation affects not only the magnitude of the available work, but also modifies how that work is distributed between locally and collectively accessible contributions. \\
Studying three observables jointly characterizes the different aspects of quantum-battery operations. Local-coherence demonstrates the redistribution of coherence at the subsystem level, global ergotropy measures the total work extractable from the joint state, and the global-local gap quantifies the additional benefit provided by collective discharge operations. The central result is that environmental placement displays an operational control parameter; its choice depends on whether the objective is total work retention, collective extraction advantage, or preservation of coherence.\\
For both the higher and lower-exchange cases, we employ the RTN-branch-resolved global master equation, with the thermal jump operators constructed from the dressed eigenstates of each conditioned Hamiltonian.  The comparison between $ J=0.45 \omega$ and $ J=2 \omega$  therefore describes the influence of exchange strength within a consistent dissipative framework.  Future work may investigate a broader range of RTN switching rates and bath spectral densities, examine the validity of full and partial secular approximations, and determine the optimal charging and discharging times.  \\
Extending the analysis to large batteries and experimentally feasible control schemes, clarifies whether the placement of environmental channels offers a scalable method for protecting and extracting useful work in noisy quantum devices. Our findings show that the spatial arrangement of the environmental channel, together with the initial coherence location, gives a practical means of controlling how coherence and energy are converted into local or collective operation in a coupled quantum battery. 

\bibliographystyle{apsrev4-2}
\bibliography{references}
\end{document}